\documentclass[twocolumn]{aastex631}

\usepackage{amsmath}

\usepackage[T1]{fontenc}

\begin{document}

\title{Re-examination of the CO absorption line in the M87 nucleus}

\author[0000-0001-8181-7511]{Norita Kawanaka}
\affiliation{National Astronomical Observatory of
Japan (NAOJ), 2-21-1, Osawa, Mitaka, Tokyo 181-8588,
Japan}
\affiliation{Department of Physics, Graduate School of Science Tokyo Metropolitan University 1-1,
Minami-Osawa, Hachioji-shi, Tokyo 192-0397}

\author[0000-0003-0292-3645]{Hiroshi Nagai}
\affiliation{National Astronomical Observatory of
Japan (NAOJ), 2-21-1, Osawa, Mitaka, Tokyo 181-8588,
Japan}
\affiliation{The Graduate University for Advanced Studies, SOKENDAI, Osawa 2-21-1, Mitaka, Tokyo 181-8588, Japan}

\author[0000-0003-0058-9719]{Yutaka Fujita}
\affiliation{Department of Physics, Graduate School of Science Tokyo Metropolitan University 1-1,
Minami-Osawa, Hachioji-shi, Tokyo 192-0397}



\begin{abstract}
We analyzed the archival ALMA data of the nuclear region of M87 and evaluate the molecular gas content from the CO(2--1) absorption line.  
We found an enigmatic variability in the absorption line depth between two epochs separated by only two months.
We reexamined the dataset used in the analysis and found that the bandpass calibration source within the same dataset also revealed a similar absorption line structure.  Furthermore, we observed a rise in the system noise temperature spectrum.  We concluded that the absorption line structure identified in a previous study, and attributed to CO(2--1), does not originate from M87 but instead results from telluric contamination, and that we still have only the upper limit on the molecular gas around the nucleus of M87. 
\end{abstract}



\section{Introduction} \label{sec:intro}

Active galactic nuclei (AGNs) are believed to be powered by gas accretion onto supermassive black holes (SMBHs) residing at the centers of galaxies.  Especially, the AGNs in massive elliptical galaxies at the centers of galaxy clusters are believed to be the heating source that prevents the development of cooling flows \citep{2000A&A...356..788C, 2007ARA&A..45..117M, 2012ARA&A..50..455F}.  Since elliptical galaxies are filled with hot gas \citep{1989ApJ...347..127F}, it has been conventionally assumed that these AGNs are fed through the spherical accretion of hot gas \citep{1952MNRAS.112..195B}.  However, recent observations show that the jet power of AGNs is not strongly correlated with the Bondi accretion rates \citep{2013MNRAS.432..530R}.  Instead, it has been shown that cold molecular gas exists in massive elliptical galaxies at the centers of nearby clusters \citep{2001MNRAS.328..762E, 2003A&A...412..657S, 2014ApJ...792...94D, 2014ApJ...785...44M, 2016MNRAS.458.3134R, 2017MNRAS.472.4024R, 2019MNRAS.490.3025R, 2016ApJ...832..148V, 2017ApJ...848..101V, 2019A&A...631A..22O, 2020MNRAS.496..364R, 2021MNRAS.503.5179N,2023PASJ...75..925F,2024ApJ...964...29F,2024MNRAS.533..771R}, which indicates that cold gas is a viable source of fuelling for AGNs.  Moreover, some theoretical and simulation studies suggest that the mass accretion onto a supermassive black hole at the center of a massive elliptical galaxy is dominated by chaotic, clumpy, and cold accretion with typical temperature of a few tens of Kelvin down to subparsec scale \citep{2005ApJ...632..821P, 2010MNRAS.408..961P, 2012MNRAS.424..728B, 2013MNRAS.432.3401G, 2015A&A...579A..62G, 2017MNRAS.466..677G, 2016ApJ...830...79M}.  In the scenario of chaotic cold accretion (CCA), the accreted gas is initially hot and X-ray emitting, and due to thermal instability the gas condensates into filaments and clouds with complex structures.  Thus, it has been widely accepted among researchers that there is generally a large amount of molecular gas generally near the centers of elliptical galaxies at the cores of galaxy clusters.

M87 is the central galaxy of the Virgo cluster, one of the closest cool-core clusters ($\sim 16.4~{\rm Mpc}$; \citealp{2010A&A...524A..71B}), and it harbors an SMBH at its center with a mass of $\simeq 6.5\times 10^9M_{\odot}$ \citep{2019ApJ...875L...6E}.  Due to its proximity and its powerful activity, including a radio jet, the AGN of M87 has been intensely studied in detail.  In particular, as in the case of other AGNs at the centers of galaxy clusters, several studies have searched for molecular gas in the nucleus of M87 \citep{1987A&A...171..378J, 1993A&A...267L..47B, 2007MNRAS.377.1795C, 2008A&A...489..101S, 2008ApJ...689..775T, 2018MNRAS.475.3004S}.  However, contrary to expectations, none of these studies has led to a clear detection of molecular gas, but only to an upper limit on its mass.

Recently, \cite{2024ApJ...974....5R} (hereafter RH24) has reported the detection of CO absorption and emission lines in the nucleus of M87 using archival data from the Atacama Large Millimeter Array (ALMA).  They analyzed the CO(1--0)\footnote{We will refer CO($J$=$m$--$n$) as CO($m$--$n$).} transition using the data under project code 2012.1.00661.S, and the CO(2--1) transition using the data under project code 2013.1.00073.S.  They reported the CO(1--0) emission showing the existence of cold molecular cloud with the estimated mass of $\sim 1.08\times 10^7M_{\odot}$ around the SMBH with a radius of $\sim 105~{\rm pc}$.  They also found the absorption lines of CO(1--0) and CO(2--1) toward the nucleus of M87, estimating the column density and temperature of the molecular cloud as $\sim 0.28\times 10^{20}~{\rm cm}^{-2}$ and $\sim 7.133~{\rm K}$, respectively.

The dataset of 2013.1.00073.S. had also been analyzed in \cite{Li2022}, in which they performed extensive atmospheric modeling, and pointed out that most of the absorption features found in the data have a corresponding atmospheric ozone transition.  They also found a calibration error that may make a too sharp absorption line in the spectrum.

In this study, we have reanalyzed one of the ALMA datasets of the M87 nucleus used in RH24 to test its interpretation.  In particular, we focus on the CO(2--1) absorption line and show that the bandpass calibration in the frequency band corresponding to the line was not performed correctly, which may have led to questionable conclusions.  In Section 2 we describe the details of the data set and the analysis. Section 3 presents the results of our analysis and the implications for the molecular gas producing the absorption line.  In Section 4 we examine the bandpass calibration of the dataset analyzed in this study, show the possibility that the dataset is not properly calibrated, and revise the constraint on the cold molecular gas around the nucleus of M87.  Section 5 is devoted to the summary.

\begin{deluxetable*}{cccccccc}
\tablecaption{Summary of the data.  (1) Date of observation. (2) Major and minor axis FWHM of the synthesized beam. (3) Position angle of the synthesized beam. (4) Native velocity resolution.  (5) Minimum/maximum baseline lengths. (6) Number of antennas.}
\tablewidth{0pt}
\tablehead{\colhead{SB Name} &  \colhead{Observation time} & \colhead{Date}$^{(1)}$ & \colhead{Beam}$^{(2)}$ & \colhead{PA}$^{(3)}$ &\colhead{$\Delta v$}$^{(4)}$ & \colhead{Baseline length}$^{(4)}$ & \colhead{NoA}$^{(5)}$\\
\colhead{} & \colhead{(sec)} & \colhead{} & \colhead{(arcsec)} & \colhead{(degree)} & \colhead{(${\rm km}~{\rm s}^{-1}$)} & \colhead{${\rm (m)}$} & \colhead{}
}
\startdata
M87\_a\_06\_TC & 876.960 & 2015-06-15 & 0.53 $\times$ 0.44 & -41.86 & 1.270 & 43.3 / 1600 & 38 \\
M87\_a\_06\_TE & 1753.920 & 2015-08-16 & 0.28 $\times$ 0.1 & 12.19 & 1.270 & 21.4 / 783.5 & 34 \\
\enddata
\end{deluxetable*}

\section{Observational Data and Analysis}
We utilize the ALMA archival data of the central region of M87, whose project code is $2013.1.00073.{\rm S}$ (Cycle 2).  The observations were done in Band 6, covering the CO(2--1) transition.  The data consists of two scheduling blocks (SBs): the first one is named as `M87\_a\_06\_TC' (hereafter, TC), and the second one is named as `M87\_a\_06\_TE' (hereafter, TE).  The information of each data is summarized in Table 1.

The data were calibrated using the appropriate versions of the
Common Astronomy Software Application (CASA) software
\citep{2007ASPC..376..127M} (ver. 4.4.0 for TC, and ver. 4.5.0 for TE).  Image deconvolution was done using CASA task \texttt{tclean} with a threshold of $3\sigma$.  We used Briggs weighting with a robust parameter of 0.5.  We adopted the velocity resolution of $3~{\rm km}~{\rm s}^{-1}$ for both datasets.  We generated the CLEANed cube data using the automasking method with \texttt{niter=100000}.  For the spectral analysis, we utilized the Cube Analysis and Rendering Tool for Astronomy (CARTA; version 4.1.0).  The rms of in the CLEANed cube data TC and TE are $\sim 6~{\rm mJy}~{\rm beam}^{-1}$ and $\sim 4~{\rm mJy}~{\rm beam}^{-1}$, respectively.

\section{Results}
In this section, we first show the results obtained from the TC and TE data. Then we discuss what is implied by these two data sets.
\subsection{TC (2015-06-15)}
The upper panel of Figure 1 shows the spectrum obtained from the TC data, extracted from the central region (where the radio continuum point source is located) of M87 with the size of $(0.77^{"}\times 0.59^{"})$ centered at R.A.: $12^h30^m49.423$, decl.: $12^d23^m28.043$.  One can see a relatively narrow feature of the CO(2--1) absorption line.  From the Gaussian fit, we obtain the continuum flux level, the flux at the line center, the line center velocity, and the line width as $1.660~{\rm Jy}$, $1. 580~{\rm Jy}$, $1228~{\rm km}~{\rm s}^{-1}$, and $11.77~{\rm km}~{\rm s}^{-1}$, respectively; the systemic velocity of M87 is $\simeq 1284~{\rm km}~{\rm s}^{-1}$ \citep{2011MNRAS.413..813C}, which was measured from the SAURON integral field stellar kinematics.

One can evaluate the optical depth of the absorption $\tau_{21}$ from the ratio of the continuum flux $S_{\rm cont}$ to the observed flux at the absorption line center $S_{\rm obs}$ using the formula,
\begin{eqnarray}
    \tau_{21} = \ln \left(\frac{S_{\rm cont}}{S_{\rm obs}} \right). \label{opticaldepth}
\end{eqnarray}
In case of TC, we have $\tau_{21,\rm TC}\simeq 0.05$.

Let us try to estimate the column density of the molecular gas producing this absorption.  In general, the column density $N$ can be calculated from the optical depth of the transition from the energy level of $l$ to that of $u$ using the formula \citep{2015PASP..127..266M},
\begin{eqnarray}
    N=\frac{8\pi\nu^3}{c^3 A_{ul}g_u}\frac{Q(T_{\rm ex})\exp(-E_l/k_B T_{\rm ex})}{1-\exp(-h\nu/k_B T_{\rm ex})}\int \tau_{ul} dv,
\end{eqnarray}
where $\nu$, $A_{ul}$, $g_u$, $T_{\rm ex}$, $Q(T)$, and $E_l$ are the rest frequency corresponding to the transition, the Einstein A coefficient of the transition, the degeneracy of the state $u$, the excitation temperature, the partition function, and the energy level of $l$, respectively.  In the case of CO(2--1), we have $\nu=230.538~{\rm GHz}$, $A_{ul}=7.36\times 10^{-7}~{\rm s}^{-1}$ \citep{1996A&AS..117..557C}, and $g_u=5$.  Following \cite{2019MNRAS.482.2934A}, we use the rigid-rotor approximation, so that $Q(T)\simeq T_{ul}/B$ and $E_l/k_B\simeq 2B$ where $B\simeq 2.766~{\rm K}$ is the rotational constant.  We cannot evaluate the excitation temperature only from the analysis of a single transition.  Here we refer to the excitation temperature obtained for the molecular cloud in the nucleus of Hydra A, $T_{\rm ex}=5~{\rm K}$ \citep{2020MNRAS.496..364R}, which is one of the few observed cases of CO absorption lines around the AGN. With the approximation of $\int \tau_{ul}dv \simeq \tau_{ul}\delta v$ where $\delta v$ is the line width ($11.77~{\rm km}~{\rm s}^{-1}$), we can estimate the column density of CO molecules as $N_{\rm CO}\sim 1.1\times 10^{15}~{\rm cm}^{-2}$.  Although the molecular clouds in Hydra A responsible for the absorption lines are supposed to be located several kpc away from the AGN, while in our case it is considered to be much closer to the AGN and the excitation temperature may be higher than $5~{\rm K}$.  However, the column density estimate would not change by an order of magnitude even if $T_{\rm ex}$ increases by a factor of few.  Using the CO-to-${\rm H}_2$ conversion factor in the Milky Way, $1.1\times 10^{-4}$ \citep{2010ApJ...721..686P}, we have the hydrogen molecule column density as $N_{\rm{H}_2}\sim 1.0\times 10^{19}~{\rm cm}^{-2}$.  This value is different from that obtained in RH24, $\sim 2.8\times 10^{19}~{\rm cm}^{-2}$.  The origin of the difference of a factor of three is not clear.

\begin{figure*}
\gridline{\fig{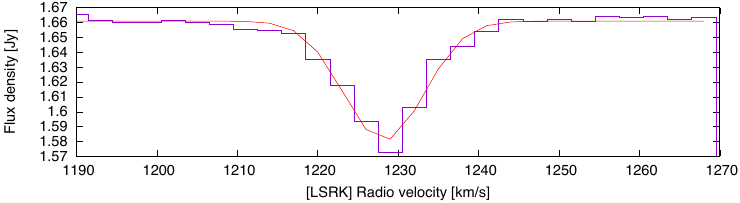}{0.8\textwidth}{}}
\gridline{\fig{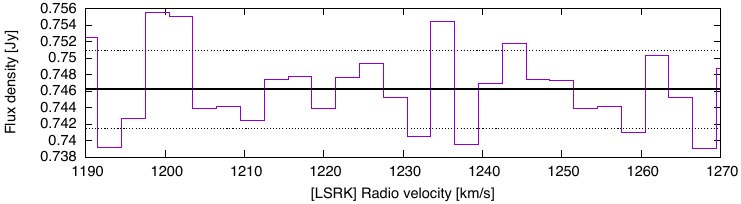}{0.8\textwidth}{}}
\vspace{-5mm}
\caption{Top: The absorption spectrum of the CO(2--1) toward the center of M87 constructed from the data TC.  The red solid line shows the fitting result (Gaussian).  Bottom: The spectrum in the same frequency range as the top panel constructed from the data TE.  The horizontal black solid line and dashed lines show the mean flux level and the root mean square, respectively.  The systemic velocity lies outside of the observed frequency range.\label{fig:spec}}
\end{figure*}

\subsection{TE (2015-08-16)}
Here we present the results obtained from the analysis of another dataset, TE, which had not been analyzed in RH24.  The lower panel of Figure 1 shows the spectrum obtained from the TE data, extracted from the same region as in the TC data.  One cannot see any absorption or emission line feature clearly.  We can evaluate the upper limit of the optical depth.  The continuum flux level is $0.746~{\rm Jy}$, and the root mean square of flux variation from the continuum level is $4.68\times 10^{-3}~{\rm Jy}$.  Adopting the $3\sigma$ level as the maximal absorption line depth expected in the data, we have the upper limit on the optical depth using Eq. (\ref{opticaldepth}) as $\tau_{21,{\rm TE}}\lesssim 0.0190$.  This corresponds to the upper limit on the column density of molecular gas, $\lesssim 4.0\times 10^{18}~{\rm cm}^{-2}$.

The continuum flux is about $\sim 45\%$ of that in the data TC.  Since such a drastic change in the continuum level only within 62 days seems unusual, we have checked the flux calibrator for each data.  For the data TC, Titan was used as the flux reference. Derived flux of the bandpass calibrator 3C 273 was $5.8~{\rm Jy}$ at $222.6~{\rm GHz}$. This shows a good agreement with the record ($5.6 \pm 0.2~{\rm Jy}$) in the ALMA calibrator catalog close in time (within 2 days).  For the data TE, the flux, bandpass, and phase calibrator are all the same, that is 3C 273.  The 3C 273 flux was $4.6~{\rm Jy}$ at $222.8~{\rm GHz}$ with a spectral index of $-0.84$, while the record in the ALMA calibrator catalog was $4.5 \pm 0.1~{\rm Jy}$ with a spectral index of $-0.75 \pm 0.04$.  We found a slight difference in the flux and spectral index between the script and catalog record, but the difference does not lead to the flux mismatch reported here.  However, there may be some unnoticed flux scaling errors which could cause such an apparent variability.  In any case, the discussions related to the absorption line would not be altered.  
\section{Discussion}
The analysis presented above shows that the CO(2--1) absorption line found in the TC data had disappeared in $\sim 62~{\rm days}$, when the TE data were taken.  To interpret this time variability, one could assume a fast moving molecular cloud in the vicinity of the SMBH or fast moving multiple continuum sources inside the AGN core.  Although it may be worth checking these scenarios, here we re-examine the data in 2013.1.00073.S.
\subsection{Bandpass Calibration}
We checked the bandpass calibration data contained in the TC and TE data.  The quasar 3C273 (J1229 + 0203) was taken as the calibrator. Figure 2 shows the spectra of the calibrator included in the TC and TE data.  One can see that the spectrum in TC has a gap at the frequencies around the CO(2--1) transition ($230.538~{\rm GHz}$).  According to the README file attached to the TC data, there was an absorption feature around this frequency range, and this feature was removed and filled by interpolation.  On the other hand, the bandpass calibrator spectrum in the TE data contains an absorption feature at the frequency corresponding to the gap found in the TC data.   This means that the bandpass calibration in the TE data was done without removing the absorption feature, which is different from what was done in the TC data.

\begin{figure*}
\gridline{\fig{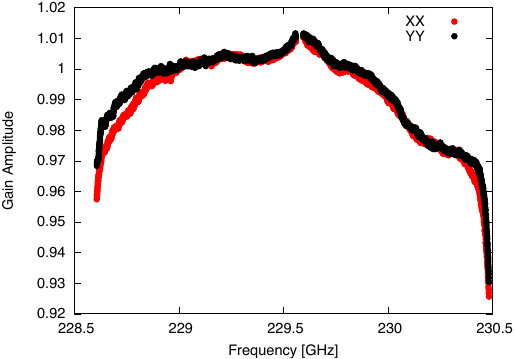}{0.8\textwidth}{}}
\gridline{\fig{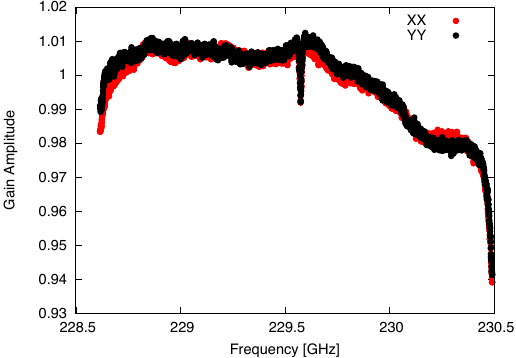}{0.8\textwidth}{}}
\vspace{-5mm}
\caption{Top: The spectra of the bandpass calibrator (3C273) averaged over all antennas in the data TC.  There is a gap at the frequency of $\sim 230~{\rm GHz}$.  Bottom: The spectra of the bandpass calibrator averaged over all antennas in the data TE.  One can see an absorption line-like feature at the frequency of $\sim 230~{\rm GHz}$.  Red and black dots represent the XX and YY polarization component, respectively.\label{fig:bandpasscal}}
\end{figure*}

To determine which calibration procedure is valid, we examined the frequency dependence of the system noise temperature ($T_{\rm sys}$) for each dataset.  Figure 3 shows the spectra of $T_{\rm sys}$ around $\sim 230~{\rm GHz}$ for the data TC.  One can see a slight increase in $T_{\rm sys}$ around the frequency of $229.6~{\rm GHz}$.  Therefore, it is likely that the absorption feature that appears in the bandpass calibrator spectrum of TE (and is supposed to have appeared in that of TC) originates from the atmospheric absorption.

\begin{figure*}
\fig{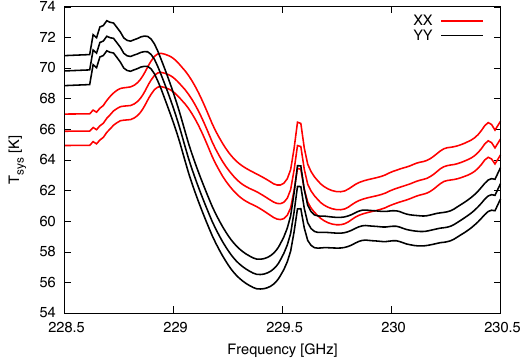}{0.7\textwidth}{}
\vspace{-5mm}
\caption{An example of the system noise temperature spectra ($T_{\rm sys}$) in the data TC.  At the frequency of $\sim 229.6~{\rm GHz}$, they have peaks.  Red and black dots represent the XX and YY polarization component, respectively.
} \label{fig:tsys}
\end{figure*}

In summary, we see that the bandpass calibration in TC was done after removing the absorption feature at the frequency around the CO(2--1) transition.  However, if this absorption feature is due to atmospheric absorption, this spectrum should be used as it is in the bandpass calibration.  The atmospheric absorption feature should appear similarly in both the spectrum of the bandpass calibrator and that of the target.  By performing the calibration using the spectrum that includes atmospheric absorption, the corresponding feature in the target spectrum should be cancelled out.  However, if the calibration is performed using the spectrum in which the atmospheric absorption has been removed, the absorption feature in the target spectrum will remain.  On the contrary, the calibration in the TE data was done taking into account this absorption feature, which seems to be relevant in this case.  To check this speculation, we reanalyzed the TC data using the bandpass calibration spectrum without removing the absorption feature that seems to originate from the telluric contamination.  Figure 4 depicts the revised spectrum obtained from the TC data with a relevant bandpass calibration, from the same region as in the analyses shown in Sec. 3.1.  One can see that the absorption feature that was found in the previous analysis (see the upper panel of Figure 1) has completely disappeared.  Having the continuum flux level and the root mean square of flux variation of $1.64~{\rm Jy}$ and $3.25\times 10^{-3}~{\rm Jy}$, respectively, we can obtain the upper limit on the optical depth using the same procedure as described in Sec. 3.2 as $\tau_{21,{\rm TC}}\lesssim 0.00596$.  This corresponds to the upper limit on the column density of $N_{{\rm H}_2}\lesssim 1.3\times 10^{18}~{\rm cm}^{-2}$. Considering that the angular size of the region examined in this analysis corresponds to $r\sim 50~{\rm pc}$, the upper limit of the molecular gas cloud can be estimated as $m_{\rm cloud}\lesssim r^2 m_{{\rm H}_2}N_{{\rm H}_2}\sim 50M_{\odot}$, assuming that the size of a molecular cloud is smaller than that of the examined region.  Even if a molecular cloud is larger than the examined region, when scaled up to the region size investigated by \cite{2024ApJ...964...29F}, $\sim 500~{\rm pc}$, it is still found to be much smaller than the mass expected from the correlation found between the molecular gas mass in the vicinity of the AGN core and the jet power \citep{2024ApJ...964...29F}.

\begin{figure*}
\gridline{\fig{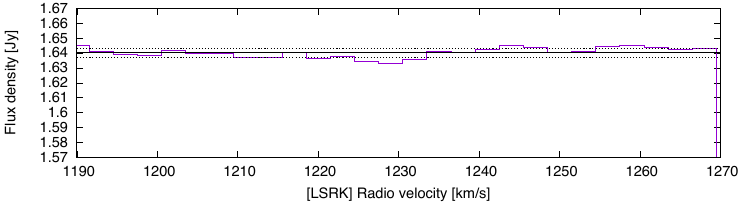}{0.8\textwidth}{}}
\vspace{-5mm}
\caption{The M87 spectrum obtained from the TC data without an artificial interpolation in bandpass calibration.  The horizontal black solid line and dashed lines show the mean flux level and the root mean square, respectively.\label{fig:spec}}
\end{figure*}

\subsection{Revised Constraint on the Molecular Gas}

From the discussion in the previous subsection, we speculate that the detection of a CO(2--1) absorption line reported in RH24 is due to a failed calibration of the data TC, which should have taken into account the atmospheric absorption.  Instead, only an upper limit on the column density of molecular gas can be obtained from the properly calibrated TC data ($\lesssim 3.4\times 10^{17}~{\rm cm}^{-2}$).  This is smaller than the value obtained from the Submillimeter Array observation at $230~{\rm GHz}$ by \cite{2008ApJ...689..775T}, $\sim 1.4\times 10^{20}~{\rm cm}^{-2}$.  Note that this value was derived from a potential emission feature, which may be due to systematic errors associated with the continuum subtraction.

RH24 analyzed CO(1--0) transition line using another dataset, 2012.1.00661.S, and evaluate its optical depth as $0.3\pm 0.02$.  Let us constrain the excitation temperature, $T_{\rm ex}$, of the molecular cloud from the optical depth ratio of CO(2--1) and CO(1--0) transitions, using the formula \citep{2003ApJ...595..167B},
\begin{eqnarray}
    \frac{\int \tau_{21}dv}{\int \tau_{10}dv}=2\frac{1-e^{-h\nu_{21}/kT_{\rm ex}}}{e^{h\nu_{10}/kT_{\rm ex}}-1},
\end{eqnarray}
where $\nu_{21}=230.538~{\rm GHz}$ and $\nu_{10}=115.269~{\rm GHz}$ are the frequencies of CO(2--1) and CO(1--0) transitions, respectively.  Assuming that the width of the CO(1--0) absorption line measured in RH24 is the same as that of the CO(2--1) line, one can constrain the optical depth ratio as
\begin{eqnarray}
    \frac{\int \tau_{21,{\rm TC}}dv}{\int \tau_{10,{\rm RH24}}dv}&\lesssim &\frac{0.00596}{0.3} \nonumber \\
    &\simeq & 0.0199,
\end{eqnarray}
which can be converted to the constraint on the excitation temperature as
\begin{eqnarray}
    T_{\rm ex}\lesssim 1.20~{\rm K}.
\end{eqnarray}
Although this is lower than the temperature of cosmic microwave background (CMB), since the excitation temperature is not necessarily equal to the thermal temperature when the collisional thermal equilibrium is not fulfilled.  Actually, there are some examples in which the excitation temperature of the molecular cloud evaluated from the ratios between CO(2--1)/CO(1--0) optical depths is lower than the CMB temperature or even negative (e.g., \citealp{2024MNRAS.533..771R}).

\section{Summary}
We re-analyzed the ALMA archival data of CO(2--1) at the nucleus of M87 to investigate the properties of the CO absorption line first reported by RH24.  In particular, we examined two datasets (TC and TE) presented in a single project and found that one of the datasets contained an absorption line feature, while the other, taken two months after, did not.  

Faced with this enigmatic results, we checked the spectra of the bandpass calibrator (3C273) in both datasets and found that the calibration was done in a different way between them: in the dataset containing an absorption line (TC) the bandpass calibrator spectrum was interpolated after removing an absorption line feature at the frequency of CO(2--1) transition, whose origin is still uncertain at this time, while in the dataset without a clear absorption feature (TE) the bandpass calibrator spectrum contains an absorption line around the CO(2--1) transition frequency.  By checking the frequency dependence of the system noise temperature and atmospheric transmission coefficient, we concluded that the absorption line removed in the former dataset is the telluric contamination, which should not have been removed to calibrate the data properly.  This failed bandpass calibration had led to incorrect interpretations of the observational data.  By reanalyzing the TC data using the bandpass calibration spectrum without removing the absorption feature, which is supposed to be the telluric contamination, one can obtain reliable results on the molecular gas around the nucleus of M87: the upper limit of its column density is $\lesssim 3.4\times 10^{17}~{\rm cm}^{-2}$, and that of its total mass is $\lesssim 13M_{\odot}$. 

In finalizing this article, we became aware of \cite{2025arXiv250303249B} on arXiv, who reported CO absorption lines (J=1--0, 2--1, and 3--2 transitions) against M87's nuclei using ALMA archive data. For the J=2--1 transition, they analyzed the same dataset (project code 2013.1.00073S) as our study, noting a faint absorption feature after atmospheric contamination removal. However, the precision of system noise temperature measurements and spectral resolution present some limitations. This raises questions about whether atmospheric absorption effects could still influence their identified CO(2--1) absorption feature in Figure D3. Regarding other transitions, while they report absorption features at consistent velocity bands in their Fig. 4, the CO(1--0) feature shows relatively modest signal-to-noise ratio, and the CO(3--2) feature appears at a slightly different velocity. These observations suggest additional confirmation may be beneficial before drawing definitive conclusions about CO absorption line detection.

\begin{acknowledgements}
We thank the anonymous referee for careful reading and many insightful comments and suggestions.  We thank Jin Koda for his helpful comments and suggestions.  We thank the EA ALMA Regional Center (EA-ARC) for their support. This work was supported by NAOJ ALMA Scientific Research Grant Code 2022-21A, and JSPS KAKENHI Grant Number JP22K03686 (N.K.), JP22H00158, JP23H04899 (Y.F.).  Y. F. is also supported by the KAKENHI Application Support Program at Tokyo Metropolitan University. This paper makes use of the ALMA data, ADS/JAO.2013.1.00073.S.  ALMA is a partnership of ESO (representing its member states), NSF (USA) and NINS (Japan), together with NRC (Canada), MOST and ASIAA (Taiwan), and KASI (Republic of Korea), in cooperation with the Republic of Chile. The JAO is operated by ESO, AUI/NRAO and NAOJ.  Data Analysis System operated by the Astronomy Data Center (ADC), National Astronomical Observatory of Japan.
\end{acknowledgements}
\bibliography{sample631}{}
\bibliographystyle{aasjournal}



\end{document}